\documentclass[slac_one]{revtex4}
\usepackage{graphicx}
\usepackage{fancyhdr}
\begin{document}
\title{{\small{2005 ALCPG \& ILC Workshops - Snowmass, U.S.A.}}\\
\vspace{12pt}
Virtual Corrections to Bremsstrahlung with Applications to Luminosity 
	Processes and Radiative Return}

\author{S. Yost, S. Majhi, and B. F. L. Ward}
\affiliation{Department of Physics, Baylor University, Waco, TX 76798, USA}

\begin{abstract}
We will describe radiative corrections to bremsstrahlung, focusing on 
applications to luminosity, fermion pair production, and radiative return
at high-energy $e^+ e^-$ colliders.  A precise calculation of the Bhabha
luminosity process was essential at SLC and LEP, and will be equally important 
in ILC physics. We will review the exact results for two-photon radiative 
corrections to Bhabha scattering which led to the precision estimates for the 
BHLUMI MC.  We will also compare the implementation of the virtual photon
correction to bremsstrahlung for fermion pair production in the ${\cal KK}$
MC to similar exact expressions developed for other purposes, and discuss
applications to radiative return in high energy $e^+ e^-$ colliders.  
\end{abstract}

\maketitle

\thispagestyle{fancy}

\section{INTRODUCTION}
In the 1990's S.\ Jadach, B.F.L.\ Ward and S.A.\ Yost calculated
the two real photon\cite{2real} and with M.\ Melles, 
the real plus virtual photon 
corrections\cite{real+virt} to the small angle Bhabha 
scattering process.  These 
corrections were used to bring the theoretical uncertainty in the luminosity
measurement, as calculated by the BHLUMI Monte Carlo (MC) program
\cite{bhlumi}, to within
a 0.06\% precision level for LEP1 parameters and 0.122\% for LEP2
parameters\cite{precision}. 

A key component of the two-photon radiative corrections calculated for 
BHLUMI was the virtual photon contribution to hard photon bremsstrahlung. 
This radiative correction is also an important contribution to the fermion
pair production process in $e^+ e^-$ annihilation implemented in the 
${\cal KK}$ MC\cite{kkmc}.  Comparisons of these results\cite{JMWY} 
to similar expressions obtained
by other authors are reviewed.  In particular, we focus 
attention on recent results for the virtual correction to hard photon 
radiation in radiative return applications\cite{KR,phokhara,epiphany2}.

\section{BHABHA LUMINOSITY PROCESS}

BHLUMI was developed into a high-precision tool for calculating
the Bhabha luminosity process at SLC and LEP, and it can continue to be 
developed to meet the requirements of a future linear collider such as the ILC. A key advantage of the program is its exact treatment of the multi-photon
emission phase space using a YFS-exponentiation procedure\cite{yfs}, so that IR
singularities are canceled exactly to all orders and the leading soft photon
effects are exponentiated, leaving only well-behaved YFS residuals to be 
calculated exactly to the order needed.  

Table 1 shows a summary of the contributions to the theoretical uncertainty
of the Bhabha luminosity process calculated by BHLUMI4.04, which includes
the complete second order leading log (${\cal O}(\alpha^2 L^2)$)
photonic radiative corrections\cite{precision}.
 The LEP1 CMS 
energy is taken to be the $Z$ mass, with an angular range between $1^\circ$
and $3^\circ$, while the LEP2 result is calculated at a CMS energy of 
176 GeV and an angular range of $3^\circ - 6^\circ$.  The portion of the 
error budget of interest here is the missing photonic ${\cal O}(\alpha^2 L)$
contribution, which is due to all two-photon radiative corrections at 
next-to-leading log (NLL) order. For ILC physics, it is desirable to reach
$0.01\%$ precision.  This is in reach for BHLUMI. Here, we will
concentrate on the photonic contributions.

The photonic part of the error estimate in Table 1 comes from comparing to an 
exact ${\cal O}(\alpha^2)$ calculation.  There are three contributions:
a double real emission term\cite{2real} which can reach 0.012\%,
a real plus virtual photon term\cite{real+virt}
 which is bounded by 0.02\%, and a two-loop
pure virtual correction\cite{precision,BVNB} making up 0.014\% in 
the LEP1 case. Adding these
in quadrature gives the error quoted in Table 1. Since all
of the exact ${\cal O}(\alpha^2)$ photonic radiative corrections are 
available, adding them to BHLUMI would remove almost all of the error
quoted for these effects.

The only remaining exact two photon contribution would then be  
``up-down interference.''  
The entire ${\cal O}(\alpha)$ up-down interference effect was 0.011\% 
at $3^\circ$ and 0.099\% at $9^\circ$\cite{updown}, and the
${\cal O}(\alpha^2)$ contribution to up-down interference would be 
supressed by an additional factor on the order of ${\alpha\over\pi}L 
\approx 0.04.$  This contribution may be neglected for small-angle
Bhabha scattering. However, a complete calculation 
of ${\cal O}(\alpha^2)$ effects,
including the full up-down interference contribution, could 
prove useful if the ILC requires a wider angular acceptance for the luminosity
monitor. Some new ${\cal O}(\alpha^2)$ computational tools and results on 
Bhabha scattering have appeared recently\cite{dixon,penin,italian,riemann,lorca}.
Comparisons to these results will be useful in gauging the precision of 
different approaches to the higher order radiative corrections in 
Bhabha scattering.

\begin{figure}[h]
\setlength{\unitlength}{1in}
\begin{picture}(6.0,2.7)
\put(0,1.5){
\begin{tabular}{|l|c|c|}
\hline \textbf{Source of Uncertainty} & \textbf{LEP1} & \textbf{LEP2} \\
\hline 
Missing Photonic ${\cal O}(\alpha^2 L)$ & 0.027\% & 0.04\% \\
Missing Photonic ${\cal O}(\alpha^3L^3)$ & 0.015\% & 0.03\% \\
Vacuum Polarization & 0.04\% & 0.10\% \\
Light Pairs & 0.03\% & 0.05\% \\
$Z$ Exchange & 0.015\% & 0.0\% \\
\hline
{\bf Total} & 0.061\% & 0.122\% \\
\hline
\end{tabular}
}
\put(-0.25,0){\parbox{3.25in}{Table 1: Summary of theoretical uncertainties for a typical 
calorimetric detectorfor LEP1 and LEP2 parameters.}}
\put(3.0,0.5){\includegraphics[width=3.0in]{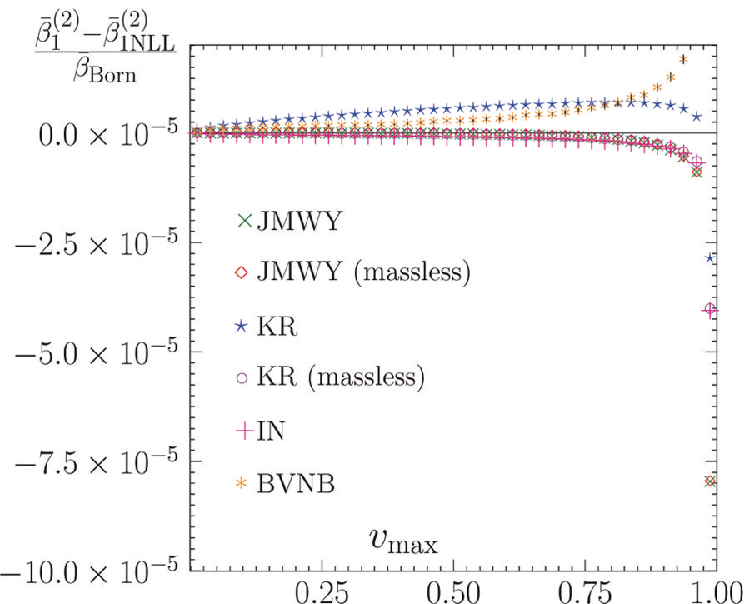}}
\put(3.0,0){\parbox{3.25in}{Figure 1: NNLL contribution to the integral of 
$\overline\beta_1^{(2)}$ for $10^8$ events as a function of the 
cut on the fraction $v_{\rm max}$.}}
\end{picture}
\end{figure}

\section{PAIR PRODUCTION AND RADIATIVE RETURN}

Another important process at electron-positron colliders is fermion pair 
production.  This process is calculated, for example, by 
the ${\cal KK}$ MC\cite{kkmc}, and
again, photonic radiative corrections are essential. 
In particular, we have presented explicit results real plus virtual
photon emission from the initial or final state fermion line. 

In the case 
of initial state radiation, emitting a single hard photon permits the final
fermion pair creation process to be investigated over a wide range of 
effective CM energies $s' = s(1-v)$, where $v$ is the energy fraction carried
away by the hard photon.  This is known as the ``radiative return'' method, 
and it can be used at a high energy collider to probe the $Z$ resonance over
a range of energies, or at a lower energy collider to measure the pion form
factor. 

Virtual photon emission is the most important radiative correction to 
radiative return.  We have compared our results for this process in the 
context of the ${\cal KK}$ MC with several other results, including 
Ref.\ \cite{IN} (IN), which is fully differential, but lacking mass corrections,
Ref.\ \cite{BVNB} (BVNB), which is differential only in $v$, but includes mass
corrections, and Ref.\ \cite{KR} (KR), which is fully differential and
includes mass corrections.  The KR result was developed for calculating
radiative return in the PHOKHARA MC, and is the newest available comparison.

We have compared these results by using them to calculate the YFS residual
$\overline\beta^{(2)}_1$, which includes the IR-finite part of single hard
bremsstrahlung including virtual photon corrections. We have shown earlier
that our result (JMWY) agrees with the IN and BVNB results analytically to 
NLL order ($({\alpha\over\pi})^2L$). We have recently shown similar analytical
agreement for the KR result at NLL order\cite{ICHEP04,compare-long}.

The NLL result is in fact very
compact, and represents the exact results to high accuracy over most of the 
range of hard photon energy fraction $v$. Without mass corrections,
\begin{equation}
\overline\beta^{(2)}_{1\ \rm NLL} = L - 1 + 3\ln(1-r_1) + 2\ln r_2 \ln(1-r_1) 
 - \ln^2(1-r_1) + 2\;{\rm Li}_2(r_1) + {r_1(1-r_1)\over 1 + (1-r_1)^2} 
 + (r_1 \leftrightarrow r_2)
\end{equation}
where $L = \ln(s/m_e^2)$, $r_i = 2p_i\cdot k/s$ with $p_i$ the incoming $e^{\pm}$ momenta and $k$ the hard photon momentum. This expression is taken as a 
baseline in comparing all of the exact expressions in a run of the 
${\cal KK}$ MC shown in Fig.\ 1.

Fig.\ 1 compares the NNLL contributions of the various exact expressions
in a MC run with $10^8$ events at a CM energy of 200 GeV for a muon
final state, using the 
YFS3ff generator (EEX3 option of ${\cal KK}$ MC). We find 
that the size of all of the NNLL contributions is less than $2\times 10^{-6}$
as a fraction of the Born cross section for $e^+ e^-\rightarrow \mu^+\mu^-$,
for photon energy cut $v_{\rm max} < 0.75$.  For $v_{\rm max} < 0.95$, all
of the results except BVNB, which is not fully differential, agree to within
$2.5\times 10^{-6}$ in units of the Born cross section.  For the final 
data point, $v_{\rm max} = 0.975$, the KR and JMWY results differ by 
$3.5\times 10^{-5}$ without mass corrections, or $5\times 10^{-5}$ with 
mass corrections. Comparisons have also been run at 1 GeV CMS energy with
comparable results\cite{epiphany2,compare-long}.

These comparisons show that we have a firm understanding of the precision
tag for an important part of the order $\alpha^2$ corrections to fermion
pair production
in precision studies of the final LEP2 data analysis and 
future ILC physics. 

\begin{acknowledgments}
Work supported by Department of Energy contract DE-FG02-05ER41399.
S. Yost thanks the organizers for an invitation to 
present this work at Loopfest IV. 
\end{acknowledgments}

\end{document}